\documentclass[doublecol]{epl2}

\usepackage{amsfonts}
\usepackage{amsmath}
\usepackage{amssymb}
\usepackage{booktabs}
\usepackage{tabularx}

\DeclareMathOperator{\Tr}{Tr}

\title{Effects of quantum deformation on the integer quantum Hall effect}

\author{
  Fabiano M. Andrade\inst{1,2} \and
  Edilberto O. Silva\inst{3} \and
  Denise Assafr\~{a}o\inst{4} \and
  Cleverson Filgueiras\inst{5}  
}
\shortauthor{Andrade \etal}

\institute{
  \inst{1}
  Department of Computer Science
  and Department of Physics and Astronomy,
  University College London,
  WC1E 6BT London, United Kingdom
  \\
  \inst{2}
  Departamento de Matem\'{a}tica e Estat\'{i}stica,
  Universidade Estadual de Ponta Grossa -
  84030-900 Ponta Grossa, Paran\'{a}, Brazil
  \\
  \inst{3}
  Departamento de F\'{i}sica,
  Universidade Federal do Maranh\~{a}o,
  65085-580 S\~{a}o Lu\'{i}s, Maranh\~{a}o, Brazil
  \\
  \inst{4}
  Departamento de F\'{i}sica,
  Universidade Federal do Esp\'{i}rito Santo,
  29075-910 Vit\'{o}ria, Esp\'{i}rito Santo, Brazil
  \\
  \inst{5}
   Departamento de F\'{i}sica,
  Universidade Federal de Lavras, Caixa Postal 3037,
  37200-000 Lavras, Minas Gerais, Brazil
}

\abstract{
In this work an application of the $\kappa$--deformed algebra in
condensed matter physics is presented.
Starting by the $\kappa$--deformed Dirac equation we study the
relativistic generalization of the $\kappa$--deformed Landau levels as
well as the consequences of the deformation on the Hall conductivity.
By comparing the $\kappa$--deformed Landau levels in the nonrelativistic
regime with the energy levels of a two-dimensional electron gas (2DEG)
in the presence of a normal magnetic field, upper bounds for the
deformation parameter in different materials are established.
An expression for the $\kappa$--deformed Hall conductivity of a 2DEG is
obtained as well.
The expression recovers the well-known result for the usual
Hall conductivity in the limit $\varepsilon=\kappa^{-1}\to 0$.
The deformation parameter breaks the Landau levels
degeneracy and due to this, it is observed that deformation gives
rise to new plateaus of conductivity in a such way that the plateaus
widths of the $\kappa$--deformed Hall conductivity are less than the
usual one.
By studying the temperature dependence of the $\kappa$--deformed Hall
conductivity, we show that an increase of the temperature
causes the smearing of the plateaus and a diminution of the effect of the
deformation, whilst an increase in the magnetic field enhances the
effect of the deformation.
}

\date{\today}

\begin{document}

\maketitle

The $\kappa$--Minkowski spacetime
\cite{PLB.1991.264.331,PLB.1992.293.344,PLB.1994.329.189,PLB.1994.334.348}
is an usual framework for studying the effects of quantum deformation on
the properties of physical systems in various contexts.
A natural manner to access such properties, starts from the construction
of the Dirac
\cite{PLB.1993.302.419,PLB.1993.318.613,CQG.2004.21.2179,JHEP.2004.2004.28},
Klein--Gordon and Schr\"{o}dinger equations
\cite{AoP.1995.243.90,EPJC.2003.31.129,PLB.1994.339.83,NPB.2001.102-103.161}
within the quantum deformation framework.
Some realizations in this direction have been reported with great
interest \cite{PRD.2009.80.025014,MPLA.2011.26.1103}.
In the relativistic context, the equations of motion are highly
nonlinear and an analysis up to first order in the deformation parameter is
the natural way to approach physical problems.
Indeed, this consideration becomes more evident and acceptable when the
problem is addressed in connection with some real physical system.
This same idea can be applied in nonrelativistic context where physical
systems in condensed matter can be studied.
It is well-known that the deformation parameter $\kappa$ can be interpreted
as being the Planck mass $m_{P}$ or the quantum gravity scale
\cite{PLB.2012.711.122} and it has implications to various properties of
physical systems, for instance, in the quantization of a linear scalar
field \cite{PRD.2007.76.125005}, in determining Landau levels
\cite{PLB.1994.339.87}, spin--1/2 Aharonov-Bohm problem
\cite{PLB.1995.359.339,PLB.2013.719.467}, Dirac oscillator
\cite{PLB.2014.731.327,PLB.2014.738.44}, Casimir effect
\cite{PLB.2002.529.256} and Dirac equation with anomalous magnetic
moment interaction \cite{MPLA.1995.10.1969}.
The investigation of physical systems on the $\kappa$--Minkowski
spacetime constitutes a proper environment in which noncommutative
theories may find applications.
In the context of recent investigations, we can mention the study of the
interference phenomena in quantum field theory
\cite{PRD.2016.93.045012}, the study of modified diffusion equations
defined on $\kappa$-spacetime to investigate the change in the spectral
dimension of  a such space
\cite{PRD.91.065026.2015,PRD.2015.92.045014} and the
analysis of bicovariant differential calculus
\cite{EPJC.2013.73.2472,JHEP.2015.055.2015}.

In this letter, we study the effects of quantum deformation (through the
parameter $\varepsilon=\kappa^{-1}$) on the Landau levels and on the
integer quantum Hall effect \cite{PRL.45.494.1980}.
As we shall show, starting by the $\kappa$--deformed Dirac equation we determine
the $\kappa$--deformed Landau levels and obtain that the deformation parameter
breaks the Landau levels degeneracy giving rise to new plateaus in the
Hall conductivity.
We also obtain upper bounds for the deformation parameter in different
materials and discuss the effect of the temperature in the
$\kappa$--deformed Hall conductivity.
We would like to emphasize that in this work we are using the framework
of quantum deformations as an effective theory, and not necessarily 
connecting it with quantum gravity.


One starts by considering a two--dimensional electron gas (2DEG) subject
to an axial magnetic field
parallel to the $z$ direction, $\mathbf{B}=B \mathbf{z}$.
Therefore, we are interested in the two--dimensional $\kappa$--deformed
Dirac equation.
Such an equation, obtained in \cite{PLB.2013.719.467} up to first order
in the parameter $\varepsilon$, reads
(in units such as $\hbar=c=1$)
\begin{equation}
  \left[
    \beta \boldsymbol{\gamma}\cdot\boldsymbol{\pi}
    +\beta m^{*}+\frac{\varepsilon}{2}
    \left(
      m^{*}\boldsymbol{\gamma}\cdot \boldsymbol{\pi }+
      es\boldsymbol{\sigma}\cdot \mathbf{B}
    \right)
  \right]\psi
  =\mathcal{E}\psi ,
  \label{eq:dirac}
\end{equation}
where $\boldsymbol{\pi}=\mathbf{p}-e\mathbf{A}$ is the generalized
momentum, $m^{*}$ is the effective mass and $\psi$ is a two--component
Dirac spinor.
The $s$ parameter corresponds to the two possible kinds of spinors in a
two--dimensional space, which is related to the signature of the
two--dimensional Dirac matrices \cite{PLB.1993.319.332}
\begin{equation}
  s=\frac{i}{2}  \Tr(\gamma_0 \gamma_1 \gamma_2)=\pm 1,
\end{equation}
and can be used to characterize the two possible spin states ``up'' and
``down'' \cite{PRL.1990.64.503}.
We take the representation for the two--dimensional Dirac matrices as
$\beta = \gamma_{0} =  \sigma_{3}$, $\gamma_{1} = i\sigma_{2}$,
$\gamma_2=-is\sigma_{1}$,
where $(\sigma_{1},\sigma_{2},\sigma_{3})$ are the standard Pauli matrices.
The vector potential associated with the magnetic field is
conveniently taken in the Landau gauge as
$\mathbf{A}=(-By,0,0)$.
Thus adopting the fermionic decomposition
\begin{equation}
  \psi =
  \begin{pmatrix}
    \psi _{1} \\
    \psi _{2}
  \end{pmatrix}=
  \begin{pmatrix}
    f(y)  \\
    g(y)
  \end{pmatrix}
  e^{ip_{x}x},
  \label{eq:ansatz}
\end{equation}
where $p_{x}\in \mathbb{R}$ is the eigenvalue of $\hat{p}_{x}$ operator, we
obtain a set of two coupled first order differential equations, namely,
\begin{align}
  \left(1+\frac{m^{*}\varepsilon}{2}\right)
  \left(p_{x}+eBy-s\frac{d}{dy}\right)g(y) = {}
  &
    \left(\mathcal{E}-\mathcal{E}_{+}\right) f(y)
  \label{eq:ode1}\\
  \left(1-\frac{m^{*}\varepsilon }{2}\right)
  \left( p_{x}+eBy+s\frac{d}{dy}\right)f(y) = {}
  &
    \left(\mathcal{E}-\mathcal{E}_{-}\right) g(y),
    \label{eq:ode2}
\end{align}
with $\mathcal{E}_{\pm}=\pm (m^{*}+\varepsilon e s B/2)$.


The second order equation implied by \eqref{eq:ode1} and
\eqref{eq:ode2} is seen to be
\begin{align}
  \left[
    -\frac{d^2}{dy^2}
    +(eBy+p_{x})^{2}
    \right]f(y)= {} &
    (\mathcal{E}^2-{m^{*}}^2)f(y)\nonumber \\
  &+(1-m^{*}\varepsilon)esBf(y).
  \label{eq:edosec}
\end{align}

\begin{figure}
  \centering
  \includegraphics[width=0.55\columnwidth]{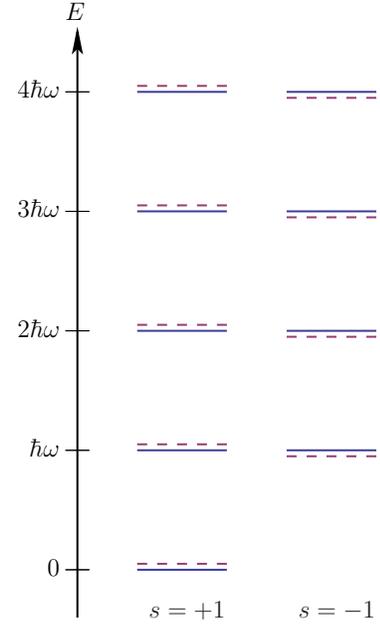}
  \caption{\label{fig:fig1} (Color online)
      Schematic representation of the Landau levels for the usual
      (solid lines) and $\kappa$--deformed (dashed lines) cases.
      In the usual case each energy level is double degenerated, while
      the ground state with $s=+1$ has zero energy.
      The inclusion of the deformation lifts the degeneracy and the
      ground state no longer has zero energy.
  }
\end{figure}

We note then that Eq. \eqref{eq:edosec} can be compared to the problem
of an one-dimensional simple harmonic oscillator.
Therefore, the $\kappa$--deformed relativistic Landau levels are
\begin{equation}
  \mathcal{E}_{n,s}^{(\varepsilon)}=\pm\sqrt{
  {m^{*}}^2+(2n+1-s)eB+m^{*}\varepsilon seB},
  \label{energy}
\end{equation}
with $n=0,1,2,\ldots$.
The above expression coincides with the result found in
\cite{PLB.1994.339.87} (for the case $p_{z}=0$) and shows that the
energy eigenvalues are modified by the deformation parameter.
Moreover, the limit of $\varepsilon\to 0$ leads us to the usual
relativistic  Landau levels \cite{Book.2012.Itzykson}, as it should be.


Let us now investigate the influence of the $\kappa$--deformed
algebra in the Hall conductivity of a 2DEG.
Thus taking the nonrelativistic limit of Eq. \eqref{eq:edosec} by setting
$\mathcal{E}^{(\varepsilon)}=m^{*}+E^{(\varepsilon)}$, with
$m^{*}\gg E^{(\varepsilon)}$, and restoring the Planck's
constant, the $\kappa$--deformed nonrelativistic Landau levels are seen
to be
\begin{equation}
  E_{n,s}^{(\varepsilon)}=
  \left(n+\frac{1}{2}\right)\frac{e \hbar B}{m^{*}}
  -(1-m^{*}\varepsilon )\frac{s e \hbar B}{2 m^{*}}.
  \label{eq:energy-nr}
\end{equation}
In the usual case (i.e., for $\varepsilon=0$), we note that the
inclusion of the electron spin causes a Zeeman splitting due to the term
${s e \hbar B}/{2m^{*}}$, thus doubling the number of Landau levels.
However, for electron with $g=2$ these levels are degenerate as a
consequence of supersimmetric characteristic of the Hall Hamiltonian \cite{EJP.19.137.1998}.
On the other hand, as we shall see, the deformation breaks this
degeneracy.
A schematic representation of the Landau levels in both cases is
depicted in Fig. \ref{fig:fig1}.

\begin{table}[t]
    \caption{
    Upper bound for the deformation parameter.
    Values used for $m^{*}$ and $g^{*}$ are from Refs.
    \cite{PRB.75.245302.2007,EL.37.464.2001}}.
  \label{tab:tab1}
  \begin{center}
    \resizebox{\columnwidth}{!}{
    \begin{tabularx}{1.1\columnwidth}{cccc}
      \toprule
        Material & $m^{*}$ & $g^{*}$ & $\varepsilon$(eV)$^{-1}$\\
        \hline
        InSb & 0.016  & -51.3  &  7.21$\times 10^{-5}$\\
        GaAs & 0.067  & -0.44  &  2.88$\times 10^{-5}$\\
        CdTe & 0.110  & -1.64  &  1.62$\times 10^{-5}$\\
      \bottomrule
    \end{tabularx}
  }
  \end{center}
\end{table}

We shall now use Eq. \eqref{eq:energy-nr} to impose an upper bound on the
deformation parameter.
In Ref. \cite{PLB.2013.719.467}, by using a relation between the
deformation parameter and the anomalous magnetic moment of the free
electron, an upper bound for the product $m^{*}\varepsilon$ was
determined as being smaller than $\alpha/2\pi\approx 0.00116$, where
$\alpha$ is the fine-structure constant.
In this manner, using the electron mass $m_{e}=5.11\times 10^{5}$ eV,
this lead us to $\varepsilon <2.27 \times  10^{-9}$ (eV)$^{-1}$.
On the other hand, it is well-known that in the realm of condensed
matter  physics, an effective mass and an effective anomalous magnetic
moment can assume different values depending on the environment and
external applied magnetic fields \cite{PR.135.1118.1964}.
For instance, in electrons confined to the GaAs layer of a
GaAs/Al$_{x}$Ga$_{1-x}$As heterojunction (which constitute a 2DEG) the
effective $g$ factor ($g^{*}$) shows an oscillatory behavior due to
exchange enhancement \cite{SS.113.295.1982}.
When a normal external field $B$ is applied to the 2DEG, and assuming
that $g^{*}$ is independent of the applied field, the
Landau levels are
$E_{n} = (n+1/2)e \hbar B/m^{*} + s g^{*} \mu_{B} B/2$, where $\mu_{B}$
is the Bohr magneton.
Thus comparing the latter with Eq. \eqref{eq:energy-nr}, we can
determine an upper bound for the deformation parameter.
In Table \ref{tab:tab1}, we show some values of upper bounds for the
deformation parameter in different materials.
Thus observing the values for the upper bound, we can conclude that
the upper bound for $\varepsilon$ depends on the material and,
for instance, it may differ in four orders of magnitude from
that of the free electron (see Table \ref{tab:tab1}).
Moreover, the product $m^{*}\varepsilon$ may assume higher values when
compared with the free electron case.
Nevertheless, Eq. \eqref{eq:dirac} is only valid up to first
order in the deformation  parameter $\varepsilon$.
Consequently, we can only use small values for $m^{*}\varepsilon$.
The value used in this work was $m^{*}\varepsilon=0.1$.

\begin{figure}
  \centering
  \includegraphics[width=\columnwidth]{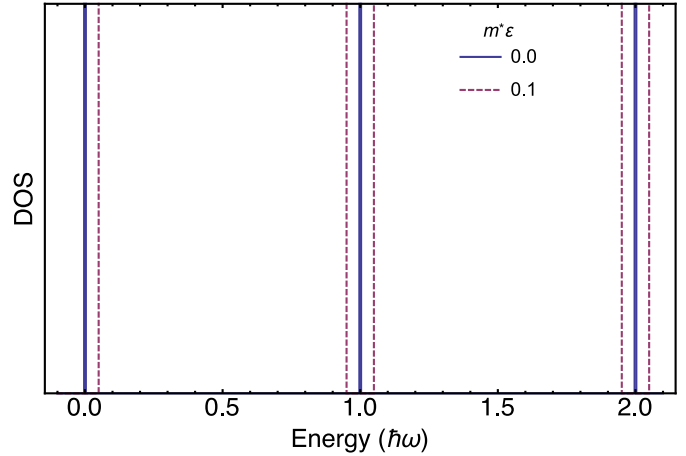}
  \caption{\label{fig:fig2} (Color online)
    The density of states versus the energy for the usual (solid
    line) and $\kappa$--deformed (dashed line) Hall system.
  }
\end{figure}

From Eq. \eqref{eq:energy-nr}, we can determine the $\kappa$--deformed density of
states (DOS) per unit of area $D^{(\varepsilon)}(E)$.
The result is seen to be
\begin{equation}
  D^{(\varepsilon)}(E)=\frac{|Be|}{2\pi \hbar}
  \sum_{n=0}^{\infty}\delta{\left(E-E_{n,s}^{(\varepsilon)}\right)}.
  \label{eq:dos}
\end{equation}
In Fig. \ref {fig:fig2}, we show the $\kappa$--deformed and usual DOS.
The discrete spectrum is schematically depicted by delta--peaks.
As we can observe, the deformation parameter $\varepsilon$ modifies the
energy spacing by lifting the degeneracy of the Landau levels with
different $s$ values, once that for $\varepsilon=0$ states with quantum
numbers $n$ and $s=-1$ have the same energy as states with quantum
numbers $n+1$ and $s=+1$.

\begin{figure}
  \centering
  \includegraphics[width=\columnwidth]{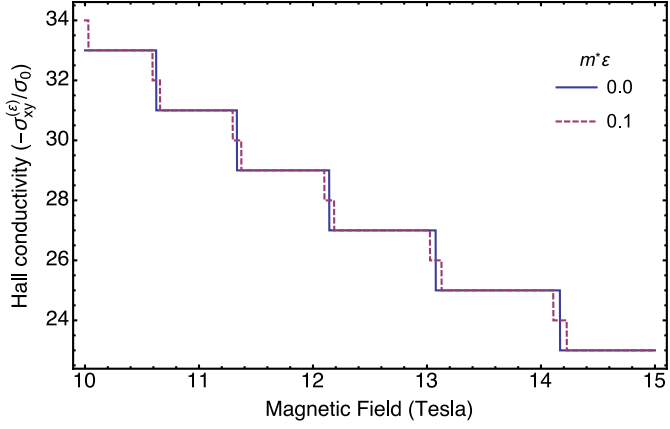}
  \caption{\label{fig:fig3} (Color online)
    The $\kappa$--deformed and usual Hall conductivity at zero
    temperature as a function of the magnetic field.
    We use $\epsilon_{F} = 4.57 \times 10^{-13}$ ergs,
    $m^{*}\epsilon_{F}/(\hbar e)=170$ T.
  }
\end{figure}

We shall now show how the lifting the degeneracy due to the deformation
$\varepsilon$, affects the quantum Hall effect by studying the plateaus
of conductivity.
In \cite{JPC.15.717.1982}, it was shown that in the linear response
approximation and when the Fermi energy is within the energy gap, the
Hall conductivity at zero temperature is given by
\begin{equation}
  \sigma_{xy}(\epsilon_{F},0)=\frac{e}{A}\frac{\partial N}{\partial B},
\end{equation}
where $A$ is the area of the surface and $N$ is the number of states
below the Fermi energy $\epsilon_{F}$.
Thus using Eq. \eqref{eq:dos} we obtain
\begin{align}
  N^{(\varepsilon)}
  = {} & A\int_{-\infty}^{\epsilon_{F}} D^{(\varepsilon)}(E) dE, \nonumber \\
  = {} & \frac{A|Be|}{2\pi\hbar}
         \left \lfloor
         \frac{m^{*}\epsilon_{F}}{\hbar e B}+\frac{1}{2}-
         \frac{(1-m^{*}\varepsilon)s}{2}
         \right\rfloor,
\end{align}
where $\lfloor x \rfloor$ is the floor function.
As the integer part in the above expression is a constant for a fixed
value of the Fermi energy, the $\kappa$--deformed Hall conductivity is
seen to be
\begin{equation}
  \frac{\sigma_{xy}^{(\varepsilon)}(\epsilon_{F},0)}{\sigma_{0}}=-
  \left\lfloor
  \frac{m^{*}\epsilon_{F}}{\hbar e B}+\frac{1}{2}-
  \frac{(1-m^{*}\varepsilon)s}{2}
  \right\rfloor,
\end{equation}
where $\sigma_{0}=e^{2}/h$ is the quantum of conductivity.
We note that the above expression recovers the well--known result for the
usual Hall conductivity in the limit $\varepsilon\to0$.
In Fig. \ref{fig:fig3}, we plot the $\kappa$--deformed and usual Hall
conductivity as a function of the magnetic field.
We can observe the presence of new plateaus (dashed lines) in
the deformed case when compared with the usual one (solid line).
These new plateaus are a direct consequence of degeneracy breaking
of the Landau levels with different $s$ values caused by the
deformation (see Fig. \ref{fig:fig2}).

\begin{figure}
  \centering
  \includegraphics[width=\columnwidth]{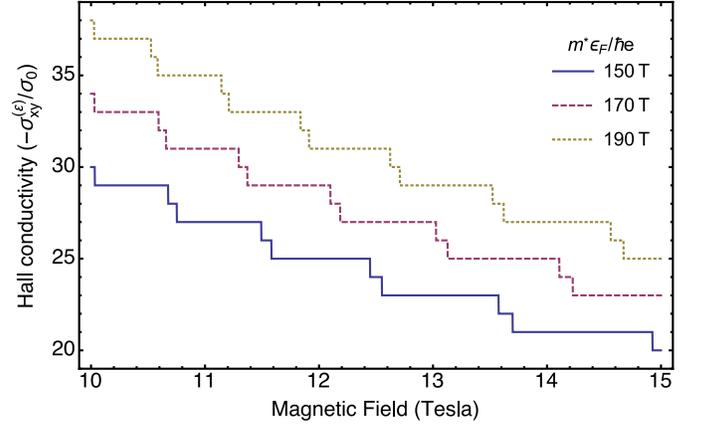}
  \caption{\label{fig:fig4}(Color online)
    The $\kappa$--deformed Hall conductivity at zero temperature as a function of
    the magnetic field for different values of the Fermi energy
    $\epsilon_{F}$ and $m^{*}\varepsilon=0.1$.
  }
\end{figure}

In Fig. \ref{fig:fig4}, we show the $\kappa$--deformed Hall conductivity for three
different values of the Fermi energy.
We can observe that increasing the Fermi energy the plateaus are
shifted to higher magnetic fields.
Moreover, the plateaus widths increase with the increasing of the
Fermi energy.
A plateau transition in the Hall conductivity occurs when a Landau level
crosses the Fermi energy, then $n_0=n+1=1,2,\ldots$ counts the number of
Landau levels occupied below the Fermi energy, i.e.,
\begin{equation}
  \epsilon_{F}= \frac{\hbar e B}{m^{*}}\left(n_{0}-\frac{1}{2}\right)
  +\frac{\hbar e B}{m^{*}} \frac{(1-m^{*}\varepsilon)s}{2},
  \label{eq:nzero}
\end{equation}
in a such way that $\sigma_{xy}^{(\varepsilon)}=-n_{0}\sigma_{0}$ on the
plateaus.
Thus considering Eq. \eqref{eq:nzero}, we can determine the plateaus
widths.
From Fig. \ref{fig:fig3} and Fig. \ref{fig:fig4}, we can observe the
presence of two plateaus widths.
The small ones are essentially due to the deformation parameter.
Their widths are seen to be
\begin{equation}
  \Delta B_{1}^{(\varepsilon)} = \frac{m^{*}\epsilon_{F}}{\hbar e}
  \frac{m^{*}\varepsilon}{(n_{0}-1)^{2}}.
  \label{eq:width1}
\end{equation}
Therefore, if $\varepsilon=0$, $\Delta B_{1}^{(\varepsilon)}=0$, and
these plateaus disappear.
Thus confirming that this plateaus are exclusively associated with the deformation.
The widths of the larger ones are given by
\begin{equation}
  \Delta B_{2}^{(\varepsilon)}= \frac{m^{*}\epsilon_{F}}{\hbar e}
  \frac{1-m^{*}\varepsilon}{(n_{0}-1/2)^{2}-(1-2m^{*}\varepsilon)/4}.
    \label{eq:width2}
\end{equation}
This latter result for the widths of the $\kappa$--deformed plateaus should be
compared with the usual ones (which can be obtained from
\eqref{eq:width2} by setting $\varepsilon=0$).
Therefore, we observe that the plateaus widths for the $\kappa$--deformed system
are less than the usual ones.
Moreover, $\Delta B_{1}^{(\varepsilon)}$ is an increasing function of
$\varepsilon$ whilst $\Delta B_{2}^{(\varepsilon)}$ is a decreasing
function of $\varepsilon$.

\begin{figure}
  \centering
  \includegraphics[width=\columnwidth]{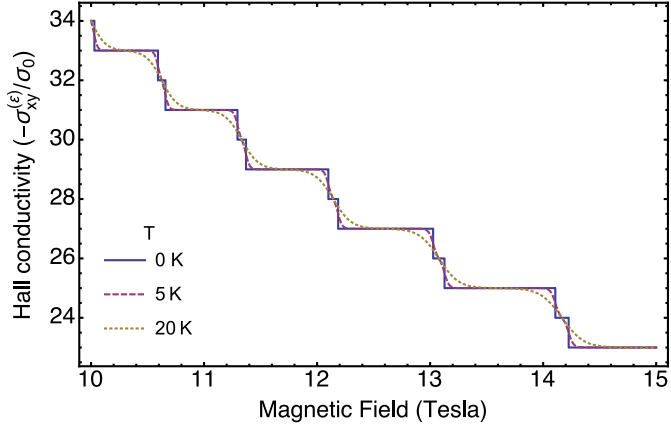}
  \caption{\label{fig:fig5}(Color online)
    The $\kappa$--deformed Hall conductivity
    $-\sigma_{xy}^{(\varepsilon)}/\sigma_{0}$ versus the magnetic
    field $B$ for three different temperatures.
    We use $\mu=4.57\times10^{-13}$ ergs.
  }
\end{figure}

\begin{figure}
  \centering
  \includegraphics[width=\columnwidth]{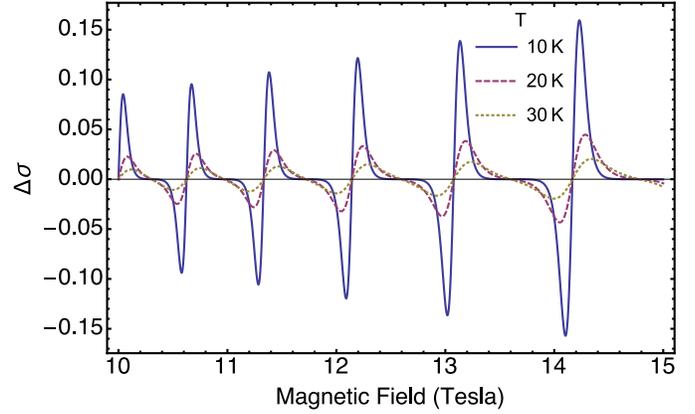}
  \caption{\label{fig:fig6}(Color online)
    The difference between the $\kappa$--deformed and usual Hall
    conductivity $\Delta \sigma$ versus the magnetic field $B$ for
    different temperatures.
    We use $\mu=4.57\times 10^{-13}$ ergs.
  }
\end{figure}

So far we have discussed the conductivity at zero temperature.
The expression for the temperature dependence of the Hall conductivity
for an ideal 2DEG, but now taking into account the two possible $s$
values and the deformation, reads
\cite{PRB.73.245411.2006,PRL.95.146801.2005,PRB.29.1939.1984}
\begin{align}
  \frac{\sigma_{xy}^{(\varepsilon)}(\mu,T)}{\sigma_{0}} = {} &
  -\sum_{n=0}^{\infty}(n+1)
  \Biggl[
  \sum_{s} n_{F}\left(E_{n,s}^{(\varepsilon)}\right)
  \nonumber \\
  &
    -  \sum_{s} n_{F}\left(E_{n+1,s}^{(\varepsilon)}\right)
  \Biggr],
    \label{eq:hallct}
\end{align}
where $n_{F}(E)=1/\{\exp{[(E-\mu)/(k_{B}T)]}+1\}$ is the Fermi
distribution function, $\mu$ is the chemical potential, $T$ is the
temperature and $k_{B}$ is the Boltzmann constant.
As shown in the Fig. \ref{fig:fig5}, an increase of the temperature causes
the smearing of the plateaus.
This smearing is even more important for the small plateaus which are
generated by the deformation.
Figure \ref{fig:fig5} also suggests that increasing the temperature, the
curve of the $\kappa$--deformed Hall conductivity comes close to the curve
of the usual one. 
In order to confirm this, let us define
\begin{equation}
\Delta \sigma =
\frac{\sigma_{xy}^{(\varepsilon)}(\mu,T)-\sigma_{xy}(\mu,T)}{\sigma_{0}},
\end{equation}
as the difference between the $\kappa$--deformed and usual Hall
conductivities.
This quantity is depicted in Fig. \ref{fig:fig6} and we can indeed
conclude that increasing the temperature, the $\kappa$--deformed Hall conductivity
approaches the usual one.
Therefore, increasing the temperature results in the weakening
of the effect of the deformation.
Moreover, by increasing the magnetic field, $\Delta \sigma$
increase.
These two behaviors can be understood in the following way.
The plateaus generated by the deformation are associated with the
breaking of the degeneracy of the states.
By increasing the temperature, the states tend to become degenerated
again.
On the other hand, the increasing of the magnetic field does the
opposite, increasing the energy difference between states with different
spin values.

\begin{figure}
  \centering
  \includegraphics[width=\columnwidth]{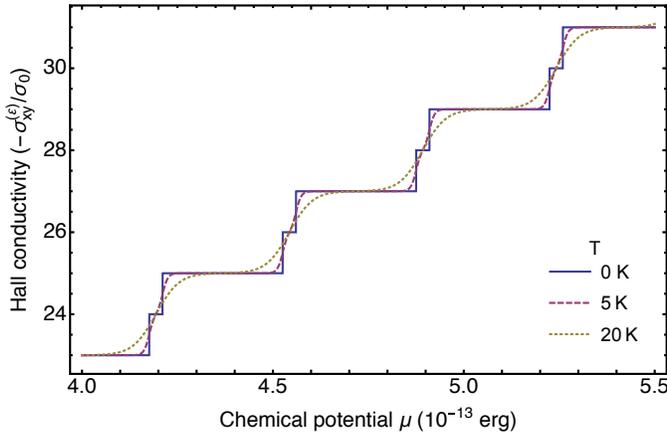}
  \caption{\label{fig:fig7}(Color online)
    The $\kappa$--deformed Hall conductivity
    $-\sigma_{xy}^{(\varepsilon)}/\sigma_{0}$ versus the
    chemical potential $\mu$ for three different temperatures.
  }
\end{figure}

For a fixed value of the external magnetic field, the expression
\eqref{eq:hallct} allows us to study the dependence of the hall
conductivity with the chemical potential $\mu$ as well.
This dependence is depicted in Fig. \ref{fig:fig7} and we can observe
that an increase in the chemical potential increases the conductivity.
Moreover, as discussed above for the dependence with the
magnetic field, an increase of the temperature results in smearing of
the plateaus.

In conclusion, we have studied the effects of the $\kappa$--deformed
algebra on the relativistic Landau levels and its consequences on the
Hall conductivity.
The $\kappa$--deformed relativistic and nonrelativistic Landau levels
were determined, and by comparing the latter with the Landau levels
of a 2DEG in a normal magnetic field, we have determined 
an upper  bound for the deformation parameter in different materials.
The values of upper bounds encountered show that the deformation
parameter is dependent on the material and may be four orders of
magnitude larger than that found for free electrons. 
It has been shown the presence of new plateaus in the $\kappa$--deformed
Hall conductivity when compared with the usual one.
These new plateaus stem from the fact that the deformation parameter
breaks the degeneracy of the Landau levels.
As a consequence of the presence of these new plateaus, the widths of
the $\kappa$--deformed Hall plateaus are less than the usual ones.
It was also shown that an increase of the Fermi energy causes the
shifting of the plateaus to higher magnetic fields.
Finally, an increase of the temperature causes the smearing of the
plateaus and decreases the effect of the deformation, whilst
an increase in the magnetic field does the opposite, enhancing the effect
of the deformation.
Given the relation between the deformation parameter and the effective
anomalous magnetic moment, we hope that some future experiment may be
able to detect the effect of the deformation on the Landau levels and
consequently on the Hall conductivity.

\acknowledgments
We thank the referees for their valuable comments.
FMA thanks Simone Severini and Sougato Bose by their
hospitality at University College London.
The authors  gratefully thank
the Conselho Nacional de Desenvolvimento
Cient\'{\i}fico e Tecnol\'{o}gico - CNPq, Brazil, for financial
support.
We also thank  CNPq for research grants
No. 460404/2014--8 (Universal) and 206224/2014--1 (PDE), 
No. 482015/2013--6 (Universal), No. 306068/2013-3 (PQ),
No. 476267/2013--7 (Universal),
FAPEMA, Brazil, Grant No. 01852/14 (PRONEM) and FAPES.

\bibliographystyle{eplbib}

\end{document}